# What is Cognitive Computing? An Architecture and State of The Art

Samaa Elnagar[1]*, Manoj A. Thomas[2], Kweku-Muata Osei-Bryson[3]
[1]Howard University, [2]University Of Sydney, [3]Virginia Commonwealth University

**Abstract**

Cognitive Computing (CC) aims to build highly cognitive machines with low computational resources that respond in real-time. However, scholarly literature shows varying research areas and various interpretations of CC. This calls for a cohesive architecture that delineates the nature of CC. We argue that if Herbert Simon considered the design science is the science of artificial, cognitive systems are the products of cognitive science or "the newest science of the artificial". Therefore, building a conceptual basis for CC is an essential step into prospective cognitive computing based systems. This paper proposes an architecture of CC through analyzing the literature on CC using a myriad of statistical analysis methods. Then, we compare the statistical analysis results with previous qualitative analysis results to confirm our findings. The study also comprehensively surveys the recent research on CC to identify the state of the art and connect the advances in varied research disciplines in CC. The study found that there are three underlaying computing paradigms, Von-Neuman, Neuromorphic Engineering and Quantum Computing, that comprehensively complement the structure of cognitive computation. The research discuss possible applications and open research directions under the CC umbrella.

## 1 INTRODUCTION

Human cognition is considered the most exceptional phenomena in all living creatures. Cognitive Computing (CC) aims mimicking human cognition to solve complex problems and embed human intelligence to machines at scale [1]. Computing systems and technologies can be classified into the categories of imperative, autonomic, and cognitive computing [2]. Imperative computing is a passive and controlled behavior system [3]. Autonomic computing is goal-driven with limited reliance on instructive and procedural information [4]. Cognitive computing embodies natural intelligence behaviors of the mind such as inference and learning [5].

CC is not only cognitive software applications but also low power and fast behaving hardware. CC hardware extends traditionally computerized systems to brain-like machines [6]. Von Neumann systems have limitations is solving combinatorics computations and real-time AI. Other paradigms such as *Neuromorphic Engineering* and *Quantum Computing* have the potential to perform combinatorics and other probabilistic calculations much faster [7]. Brynjolfsson and McAfee [8]) highlighted that organizations need cognitive systems that can handle complexity, make confidence-based predictions, learn actively and passively, act autonomously, and reflect a well-scoped purpose. Some organizations expect that CC would enhance decision making process and cut costs [9]. However, some researchers confound CC with AI but, AI is a subcomponent in the CC architecture [10]

Current AI Limitations and Objectives of CC

AI is not intended to mimic human cognition but to build the best possible algorithms to solve several problems. Current AI systems employ symbolic logic intelligence to perform a limited number of mental activities [11]. For example, cognitive visual recognition systems may mimic the human vision. Existing AI lacks robustness against changes in the input statistics and they are not robust to adversarial examples and domain adaptations [9]. In addition, current cognitive systems are confined by the architecture of the von Neumann platforms which separate memory from processing units and depends mainly on Boolean logic [12]. Russell and Cohn [13]) summarized the challenges of current AI systems as the ‘‘Moravec's paradox, i.e., computers might exhibit adult level performance on intelligence, but it is difficult or impossible to give them the skills of a one-year-old when it comes to perception and mobility.'' On the contrary, CC aims at maximizing the cognitive competences of machines while minimizing resources. CC introduces new hardware that offers more intelligence with less resources. CC aims to build scalable cognitive systems that employ human brain like processing mechanisms while maintaining reasonable consumption of resources.

### 1.1 Research problem

Research in CC is very diverse, different studies have linked articles that focus on AI, mathematics, and electronics to the body of knowledge of CC [14, 15]. This variety led to indistinct research agenda on the topic affecting many scientific fields such as Information Systems which find a disruption in the basic assumptions about how to employ CC [16]. To reach a comprehensive delineation of CC, attentive analysis of the existing literature is required. Most previous research used inductive approaches to reach a taxonomy of CC [17-19]. However, these inductive approaches followed mostly ad hoc qualitative methods that are intuitive in nature. Another issue with previous qualitative interpretations is that they lack systematically coupling the underlying elements to-from the interpretation. Thus, a theoretically grounded architecture of CC, that uses rigorous statistical techniques, is needed. This is not to say that the results of qualitative approaches are to be discounted. On the contrary, they should support the statistical findings to cement the explicit and implicit meanings within the data [20]. In other words, qualitative research results could be considered as a preliminary analysis or an initial premises that need to be further verified with quantitative analysis.

This paper aims to investigate the nomothetic nature of CC and proposes an architecture that explains its teleological structure (explaining the purpose COCS serve). This investigation uses different quantitative techniques to verify the nature of CC based on the categorization of published literature. The construction of a unified architecture will shed light on how diverse research areas under the CC umbrella may come together to address emergent areas of related research. We believe Cognitive Computing has two new pilar paradigms that complement the limitation in current cognitive systems which are *Neuromorphic Engineering and Quantum Computing* [21]. The study thus comprehensively explores the state of the art in CC development and recommends future research directions.

## 2 SCIENTIFIC FOUNDATIONS OF COGNITIVE COMPUTING

The early foundation of CC was introduced in the early 1960s when Simon [22] published the book - “*Cognitive Science,* the new science of the artificial”. Cognitive Science aims to analyze human intelligence in information processing context and sets intelligence as its main objective. As Simon mentioned “Artificial Intelligence and Cognitive Psychology are two channels of communication that were largely confined to their separate disciplines. Only with the establishment of *Cognitive Science,* a channel was created that cut squarely across the disciplinary boundaries”.

On the same venue, Intelligence Science is an interdisciplinary science dedicated to joint research on basic theory and technology of intelligence. Intelligence science cares about applying abstract intelligence to machines [23]. Intelligence is intangible and is not a substance that can be processed but Intelligence are *symbols* and it can be represented in symbolic structures. Intelligence feeds on knowledge to survive [24]. Newell [25] pointed out that symbol-level systems and knowledge-level systems are the necessary general structures to obtain general intelligent behavior.

According to Pinker [26], to understand the human mind, it is necessary to understand the principles of many natural sciences such as *Cognitive Neuroscience, Cognitive Psychology, and Brain Science*. However, to apply intelligence to artificial systems, it is necessary to integrate *Computer Science* and *Information Science* with natural sciences to create applied intelligence science such as *Cognitive Science and Intelligence Science* as shown in Fig. 1.

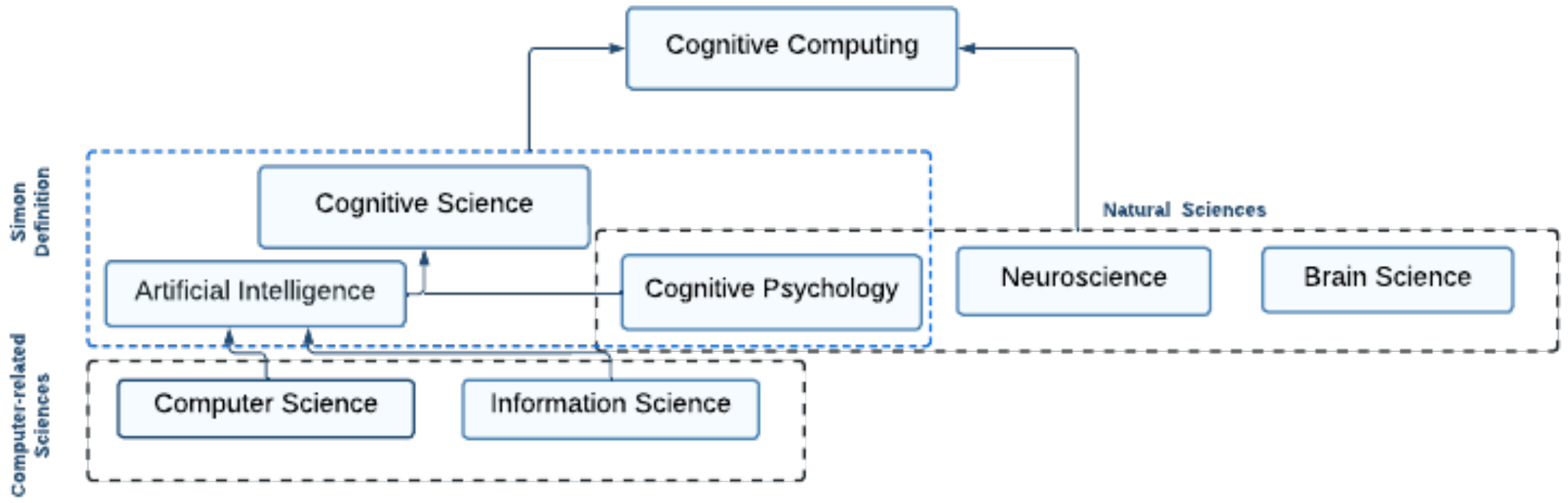

Fig. 1: The Scientific Basis of Cognitive Computing.

## 3 METHODOLOGY

Since there are several distinct research disciplines in cognitive computing, we searched beyond the current technologies listed as CC, to find out how these different research disciplines contribute towards CC. To reach saturated evidence about the architecture of CC, we employed a combination of inductive and deductive approaches using different statistical methods as discussed in the following sections. Then, the findings of this research are compared with the findings of the previous research. The analysis performed in the study is an extension of the preliminary thematic analysis conducted in [27]. However, the focus of this research is to conduct further statistical analysis to reinforce or undermine the findings of the previous qualitative analysis.

### 3.1 Prior Research

[27] performed a systematic literature review (SLR) analysis on CC using qualitative analysis methods and a *Phenomenological Reflexives* approach [28]. We reused the output references of their SLR. Then, we conducted the same four search rounds to search for newer references. In the first round we used the same keywords. However, some research claimed quantum computing as one of paradigms under CC. So, in the second round, we added “quantum cognition, quantum cognitive computing” to search keywords. We recalled the same forward and backward search and the same inclusion and exclusion criteria to filter articles, but we also included Quantum hardware along with Neuromorphic hardware in the search processes.

### 3.2 Search Process Findings

By accumulating and synthesizing evidence from published literature, The qualitative analysis was similarly performed using Nvivo 12 for thematic analysis. The thematic analysis was consistent with Elnagar findings and there are four distinct, yet integrated research lines of CC related to the basics of intelligence, hardware that mimics the brain (brain-like hardware), software algorithms, and cognitive systems.

## 4 STATISTICAL ANALYSIS OF CC LITERATURE

To categorize varied topics in CC research and converge them into cohesively related themes. We followed the scientific methodology developed by Nickerson, Varshney and Muntermann [29]) for developing literature taxonomy.

### 4.1 Topic Modeling Analysis

The first statistical analysis performed is Topic modeling (TM) to identify the themes or topics in the CC literature [31].

### 4.2 Topic Modeling Experiments

The experiments were conducted on the Google Collaboratory cloud platform. The entire modeling process could be divided into four main phases: 1) corpus preprocessing, 2) corpus formatting, 3) topic modeling and 4) relationships extraction as shown in Fig. 2.

### 4.3 Corpus Preprocessing

We followed Brust, Breidbach, Antons and Salge [41]) to pre-process the text corpus for analysis. The textual data corpus is synthesized from the abstracts of the of the systematic literature review papers. First, a python code was developed to automatically select and save the manuscript abstracts to an excel file for ease of analysis. However, we converted the research files from pdf to MS word format to avoid errors in reading pdfs. Then, each file's abstract is stored in one excel cell

The analysis of results also found that LDA and Bi-LDA using Sklearn gave more interpretable topics than LSI, and HDP. Results of the first round are shown in (Fig. A1.1, A1.2, A1.3) in the Appendix. Another observation was that the results of IBM topic models took longer time and turned out irrelevant because the topic models is shared and trained by many users' domains. Therefore, IBM topic models were excluded from the next iteration of experiments.

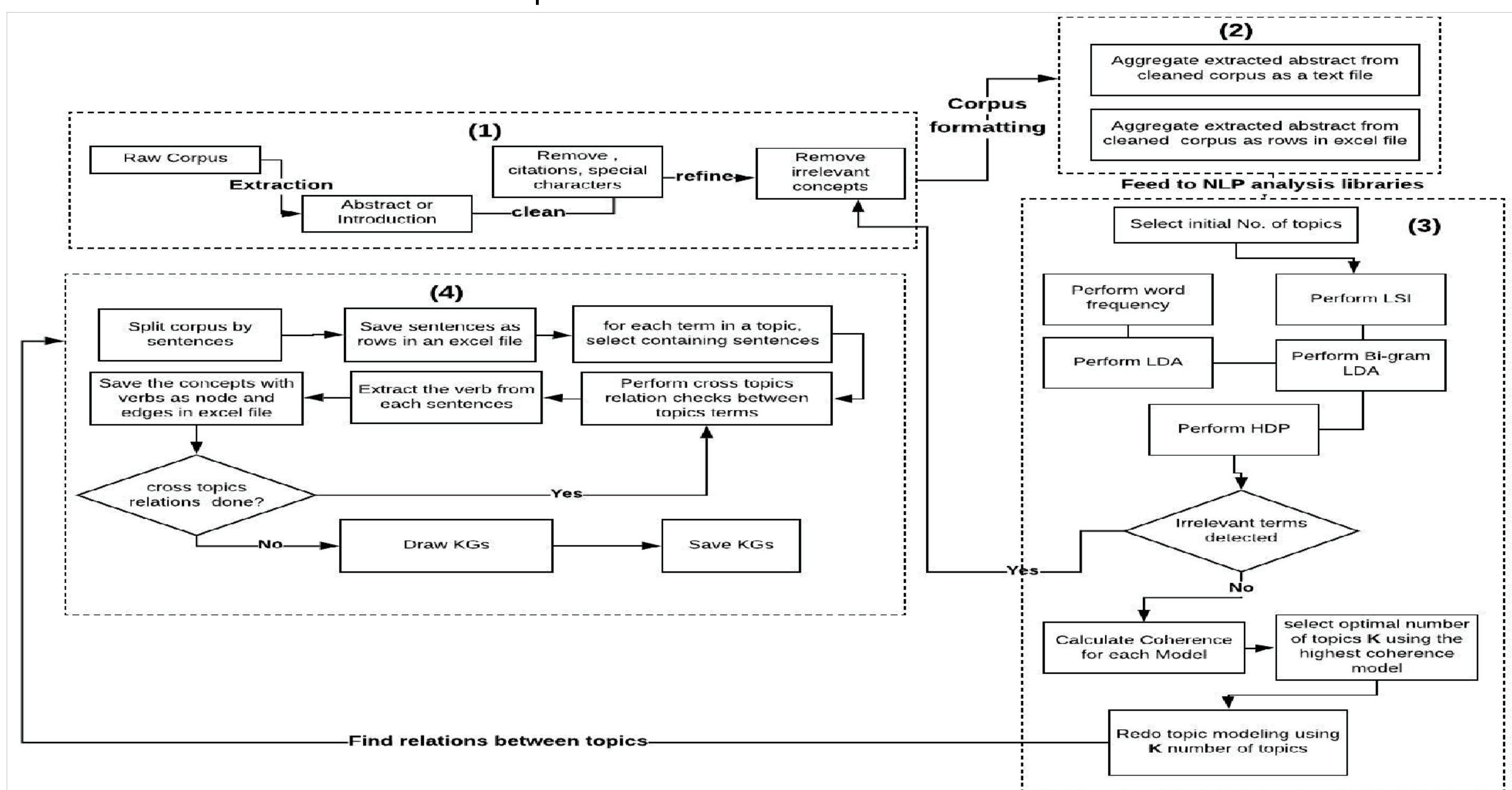

Fig.2: Topic Modeling Experiment Phases

---

### 4.7 Model Interpretations

Interpreting the *pyLDAvis* Fig.s below, a coherent representative topic model will have fairly larger size - the larger the topic circle, the more prevalent is that topic [45]. In addition, non-overlapping topics refers to their distinctness. The greater the inter-topic distance, the more representative is the model. The four topics shown in Fig. 6.c are exclusively located in the four quarters and they are well separated. The size of each topic is fairly large. Concepts within topics are human friendly and representative of the terms found in the corpus. Analysis indicates that LDA generally performed better than other TM approaches (HDP, LSI). Generally, LDA performs well when the underlying topics are geometrically concentrated and wellseparated or when documents are associated with small sub-sets of topics [36].

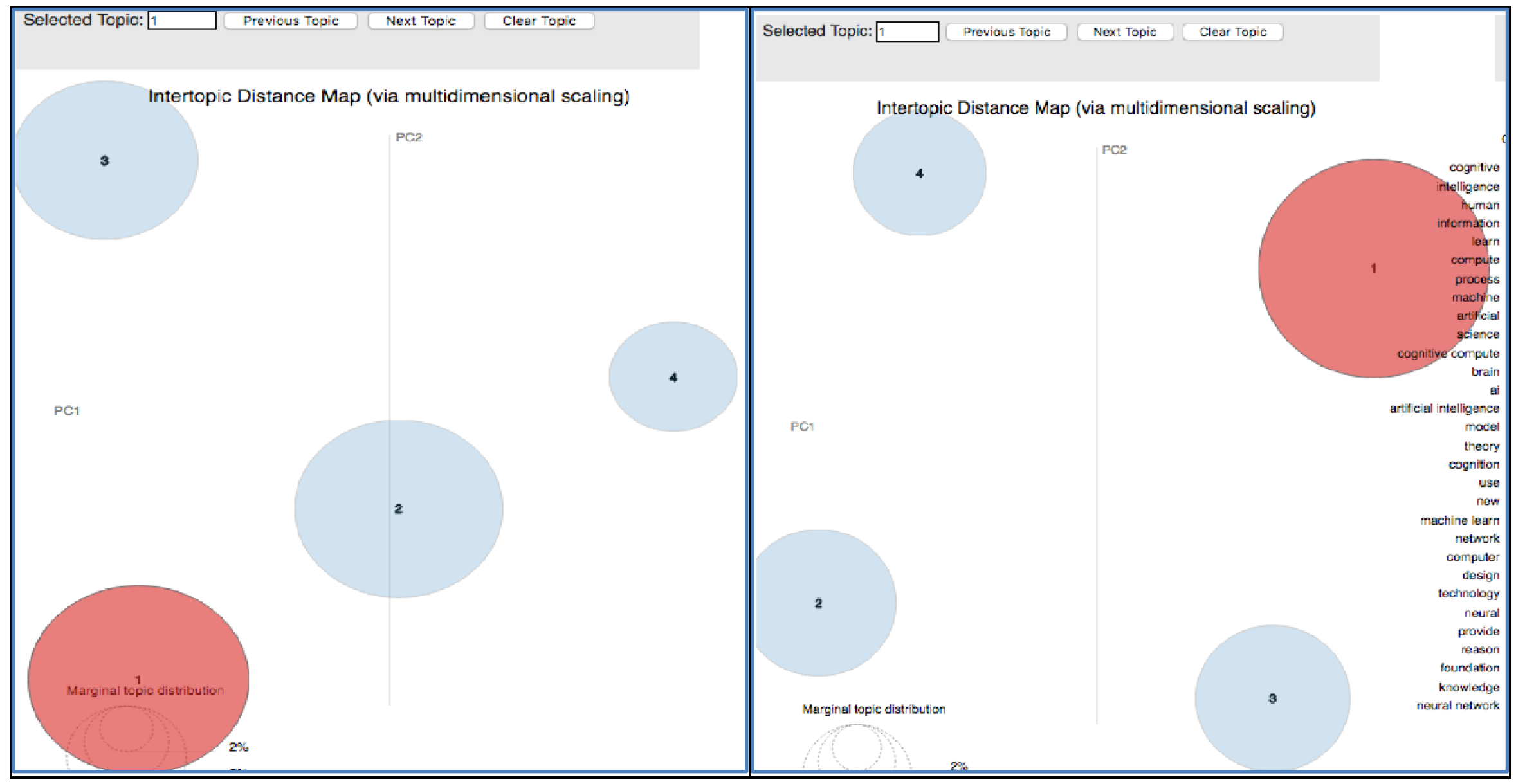

Fig. 6.a. Visualized LDA model using pyLDAvis libraries at K=4 using Genism

Fig. 6.b. Visualized LDA model using pyLDAvis libraries K=4 using Sklearn

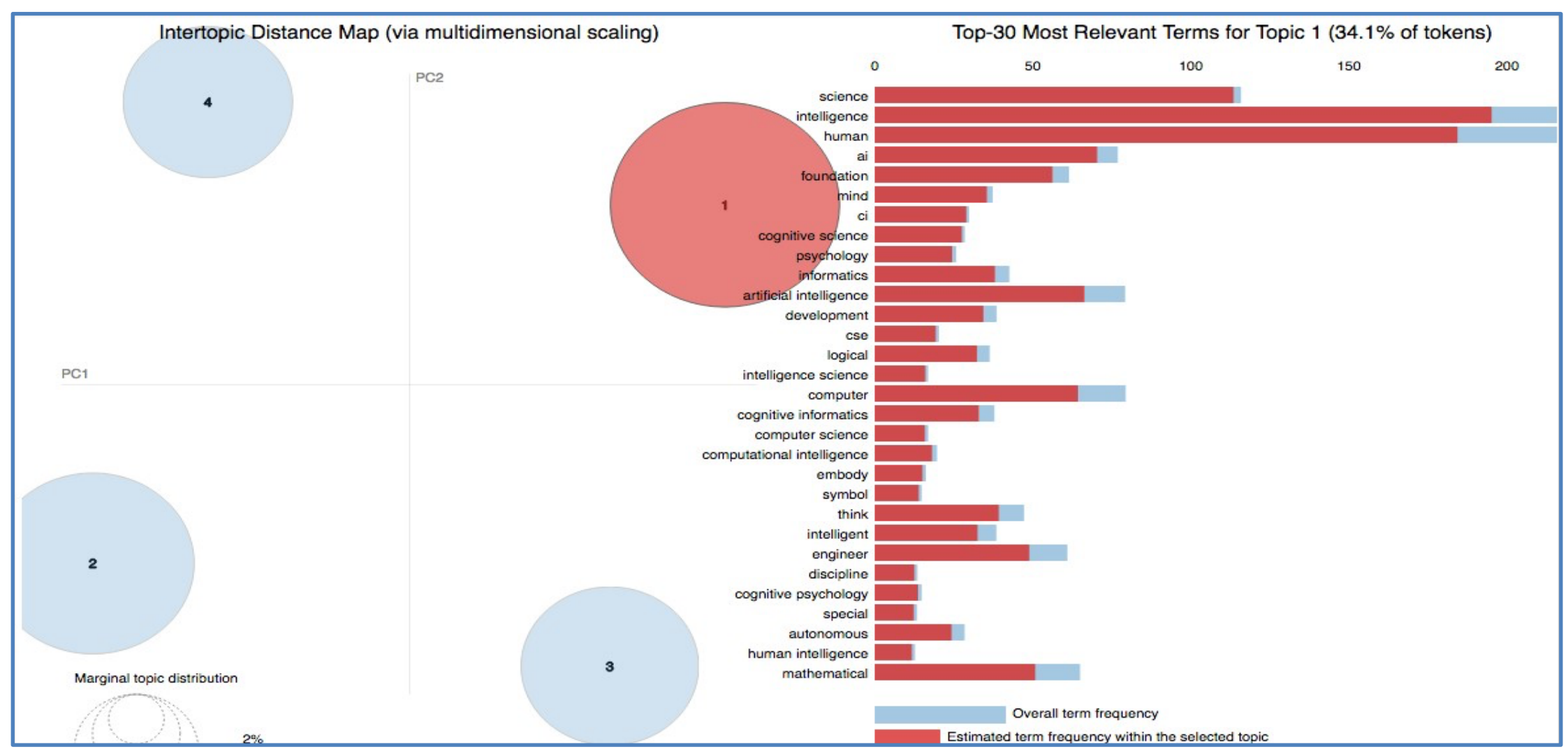

Fig. 6.c. Visualized BI-LDA model using Sklearn and pyLDAvis libraries at $\lambda$ =0.33 and K=4.

### 4.3 Comparing Statistical Analysis with Qualitative Analysis

The main four topics resulted from the analysis are shown in Table 3. In each topic, the most expressive 10 concepts according to their saliency and relevancy are listed. However, to further check the rationality of the TM findings, previous qualitative research findings are also presented to compare and contrast the findings.

If qualitative analysis themes are consistent with statistical topic modeling results, this will reinforce the proposed CC architecture. Each topic represents a distinct research area. To measure the similarity between concepts generated from the topic modeling experiments and the themes coded from the thematic analysis, we used the popular *Jaccard similarity* measure to calculate the similarity between the statistical analysis findings and qualitative analysis finings. *Jaccard similarity* calculates the ratio between the intersection and union of the two sets of items.

*4.3.1 Relationships between Concepts*

The TM process identified four foundational topics of the CC architecture, where each topic is presented by the most salient 10 concepts. If there are four diverse topics that represent CC that are verified by both statistical and qualitative analysis, there should be a relation or dependency between these topics. One of the approaches that gained popularity in find relationships between concepts is Knowledge Graphs (KG). A knowledge graph is a structured representation of facts, consisting of entities, relationships and semantic descriptions [49].

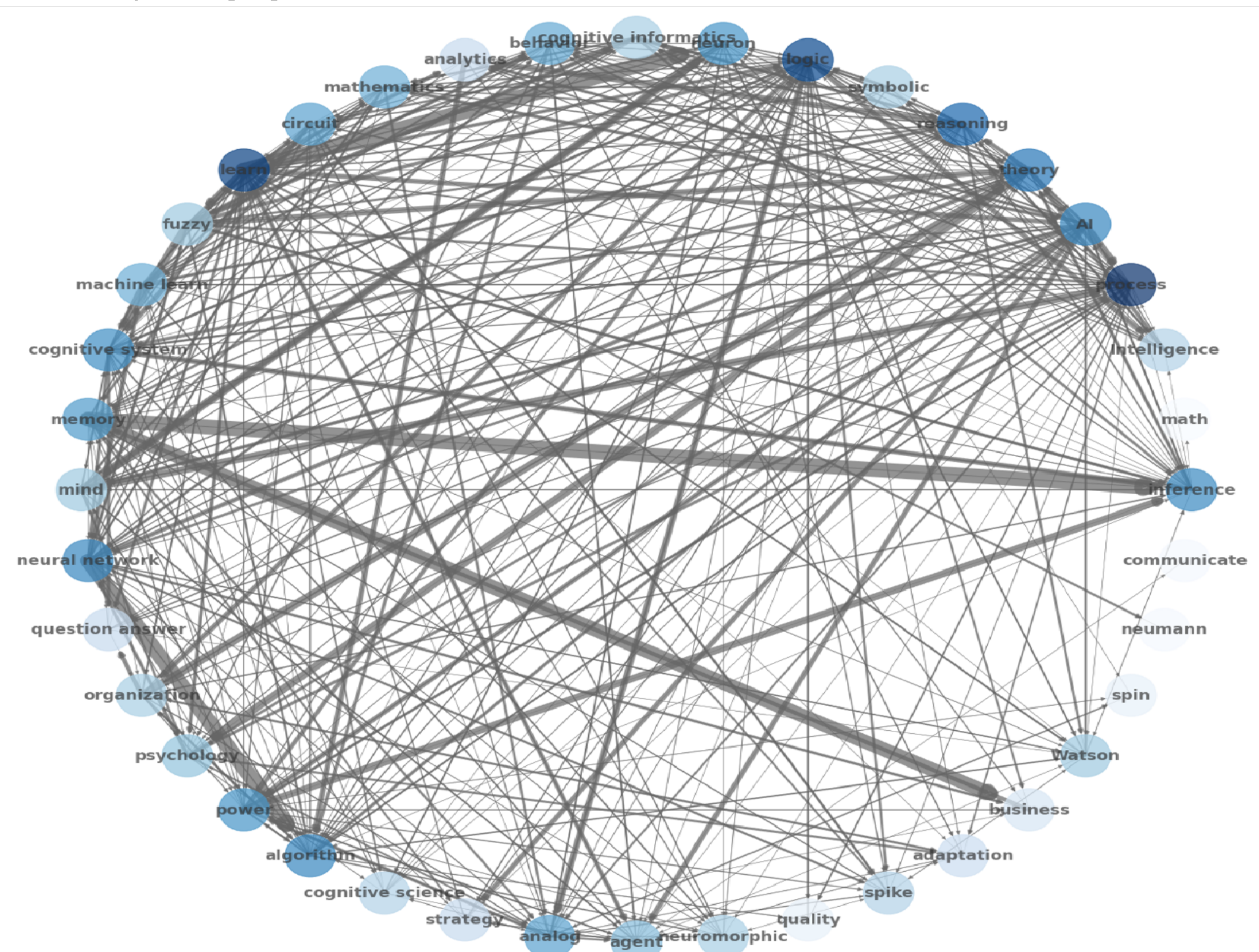

Fig. 4: Knowledge Graphs Representing Relationships between Concepts Based on Their Weight (Circular Mode).

Performing cross topic relationships resulted in more than eight thousand relationships. This was unreadable when visualized in a graph. Notably, most nodes and edges pairs (pair of concepts) are repeated. Therefore, we reduced the edge list by creating a *weight* attribute which represents how often a pair of concepts appeared in the edge list. The weight is visualized as the thickness of the line between a node and an edge - the thicker the relation line, the stronger is the relation between two concepts. The reduction algorithm resulted in 613 pairs of relationships along with their *weight.* The reduction process is represented in Algorithm 3 and the resulting knowledge graph is visualized in Fig. 7. The figure was plotted using *networkx*, a python library for plotting knowledge graphs in the form of nodes and edges [52]. As shown in the Figure., there are a total of 40 concepts (four topics with 10 concepts each).

*4.8.3 Foundational Topics of CC architecture*

By summarizing the concepts in Topic 0, we can observe that they characterize human cognitive functions and mental models in mathematical symbolic format. Topic 0 captures the core scientific knowledge archived in mathematical forms [54]. In most cases, the cognitive functions are derived from brain sciences such as neuroscience to form mathematical versions of brain models [55]. The mental models adopt a spatiotemporal computation approach [56], and the quantum probabilities to handle uncertainty beyond the limitations of the binary and bounded probabilities [57].

## 5 COGNITIVE COMPUTING ARCHITECTURE

Based on the analysis findings, we can conclude that the structural architecture of Cognitive Computing could be divided into four dependent sections: *Representation of Intelligence, Brain-like hardware, Cognitive Algorithms and Software, and Cognitive Systems .* The pyramid structure depicts the *dependency* of each section on the previous sections in the pyramid. So, the development in any of the four sections is counting towards the development in CC. To develop a cognitive system, a bottom-up approach should be followed using the pyramid-like structure. The architecture represents a blueprint that embodies the logic, hardware and software needed for learning, adaptation, and evolution of cognitive systems.

At first, intelligence must be encoded into symbolic form as a universal representation of the human mental activities to be easily applied to hardware and software. This section of the CC research is committed to the *Representation of Intelligence*. The *Brain-like hardware* section is representing the electronics that mimic the nervous system of the brain in terms of memory, processing, and communication channels. This new hardware also requires new programming languages to build software that run on the *Brain-like hardware*. At the top of the pyramid, *Cognitive Systems* are developed using the underling software, hardware and logic. Thus, research in CC integrates these four different yet related research areas. It is important for researchers to understand the big picture and connect the four research areas while conducting research on cognitive computing. In the following section, the state of the art in CC research is discussed in detail.

## 6 COGNITIVE COMPUTING: STATE OF THE ART

In this section, we present the state of the art in CC research by tracking the progress in each of the four sections of the CC architecture. We follow a bottom-up approach to present how the advances in each of the four sections helped the development of the sections above. Thus, research on cognitive systems at the top of the pyramid is dependent upon the research progress in the underlying sections. Next, the state of the art in each section is presented in detail.

### 6.1 Representation of Intelligence

At the foundational level of CC pyramid architecture, the representation of intelligence as mathematical format ensures consistent performance across different hardware or software platforms. Among the two categories of applied mathematics, i.e., analytical mathematics and denotational mathematics, the latter is more suited for representation of intelligence [65]. Denotational mathematics (DM) is a form of abstract algebra that aims at formalizing mathematical objects (called denotations) to describe the meanings of language expressions and system behavior with abstract concepts and dynamic processes [65]. Notable progress has been achieved in denotational mathematics such as the development of *Concept Algebra, System Algebra, Real-Time Process Algebra (RTPA*), *Visual Semantic Algebra (VSA), And Inference Algebra* [66]. Without DM, machines reuse developed knowledge to identify similarities in new problems nor cognition identify others delineations of sentences [67].

based on the notion that human brain is the source and final destination of information [78].

### 6.2 Brain Hardware

Despite the fact that many cognitive systems are classical Von Neumann computers, they suffer many challenges such as “intellectual bottleneck”, deficiency in real time decision, high power consumption, and slow performance [6]. Traditional computers separate memory from computation, requiring information to shuffle between the memory and the CPU via a bus which slows down processing. In addition, the processing power needed increases as the communication rate (clock frequency) increases [79]. This led to calls for brain-like information processing hardware that consume low power, respond at real time, and can be easily scaled. So, this hardware could be widely embedded to robotics and Internet of Things (IoT) [80].

### 6.3 Cognitive Algorithms and Software

Cognitive Computing is expected to follow all the three schools of artificial intelligence: *connectionism*, *symbolism*, and *behaviorism.* [24]. AI in CC is distinguished by attribute-efficient learning algorithms and tractable learning.Traditional Neural Network (NN) is limited to understand syntax and symbols. While *Deep Neural Networks (DNNs)* have significantly expanded the inference and unsupervised learning capabilities of machine learning techniques, they are computationally complex and exhaustive [99]. Recently, more research is dedicated to developing AI algorithms to *Neuromorphic* and *Quantum* hardware.

### 6.4 Cognitive Systems

Cognitive systems are at the top level of the pyramid-like architecture of CC. Cognitive systems could be in the form of applications, agents, services, and APIs that align with the pyramid like architecture including representation of intelligence, brain like-hardware and software algorithms. The hallmark of a cognitive system is its abilities to function effectively in circumstances that were not planned explicitly when the system was designed [16, 115]. According to Pylyshyn [116]) cognitive systems take two approaches: the first is the symbolic information processing systems approach which uses cognitive codes to determine the operations of the system. The second combines connectionist, dynamical, and enactive systems under the general approach of emergent systems.

To summarize the state-of-the-art findings, there are three paradigms that comprehend Cognitive Computing systems: Von-Neumann systems, Neuromorphic systems, and Quantum systems. Each paradigm has its scope of applications, strength, and weaknesses. A comparison between the three paradigm is presented in Table 4.

## 7 DISCUSSION

Previous research defining what is CC depended merely on researchers' opinions without apt justification. The main contribution of this study is to explore the nature of CC by statistically analyzing the CC literature and comparing it with previous research findings. The study also proposed an architecture for CC as shown in Fig. 9 and discussed the state of the art in CC research. There is main three computing paradigms that encompass the complex human mind cognitive functions which are: *Von-Neumann computing, Neuromorphic computing, and Quantum computing*. Each paradigm has its own strengths and applications under the CC umbrella.

Table 4. A comparison between Von-Neumann systems, Neuromorphic systems, and Quantum systems.

| **Aspect** | **Von-Neumann systems** | **Neuromorphic systems** | **Quantum systems** |
|---|---|---|---|
| Signal Used | Digital | Analog/Digital | Analog/Digital |
| Data Format | Bits | Spikes | Qubits |
| Conductors | Electrons | Artificial neurons and synapses | Trapped ions/ Photons/ Molecules |
| Memory | Separate | Co-located | Separate / Co-located |
| Application Focus | Mathematical computations | AI/ Neural networks | Combinatorics, optimization, AI |
| Strength | Stable, low cost, widespectrum applications | Simple Parallelization, Scalability, Low Power, Stochasticity, Graph AI | Quantum Parallelism, Optimization speed, Stochasticity |
| Stand-Alone systems | Yes | No | Yes |
| Scalability | Vertical/ Horizontal | Horizontal | Horizontal |

| Weaknesses | High-resource consumption, Slow | Narrow-focused, Complexity | Costly, Narrow-focused, Decoherence |
| --- | --- | --- | --- |
| Purpose | Multi-Purpose | Narrow-Focused | Narrow-Focused |
| Computation trigger | Data Driven | Event Driven | Both |

The pyramid architecture for CC can be summarized as: *a Representation of Intelligence layer* that is responsible for encoding complex cognitive functions. Different cognitive functions require different types of representations. So, cognitive cognition functions require encoding different that Neuro related ones. The next layer is the *Brain-Like Hardware layer* at which different types of hardware are necessary to run the special encodings of intelligence. The following layer discusses the special programming languages, AI and software algorithms suitable for the brain-like hardware. So, the third layer that aims to develop the software that could communicate with both the underlying hardware and classical Von-Neumann Systems. At the top level of the architecture is the cognitive systems layer that are built using the layers below.

To summarize the state of the art of CC, research has taken three directions depending on the hardware used. Currently, most of the cognitive systems use conventional Von-Neumann hardware with conventional sequential programming. The other direction is using NE-based-electronics that consist of neurosynaptic cores and communicate using spikes. NE cores need special programming languages such as Corelet programming and new AI algorithms such as spike-based deep learning algorithms. NE based cognitive systems are expected to use both symbolism and connectionist intelligence. The third is Quantum Computing paradigm to solve the deficiency in decision making and uncertainty in solving complex probabilistic problems.

Thus, the ultimate goal of CC is to develop cognitive systems that mimic the cognitive abilities of humans. However, research in representation of intelligence, hardware, or software that contributes to building cognitive systems is considered a CC research contribution. Although cognitive systems based on NE and Quantum circuits are still under research, they are expected to surpass the conventional hardware in terms of adaptation, cognition, speed, memory, and power consumption [125].

To help unify research goals, we recommend that NE and Quantum researchers include CC as a main keyword in developing NE or Quantum hardware/software. Our analysis found that most research in NE or QC do not include CC as keyword. Nevertheless, we believe that the next decade we will find notable improvements in *Cognitive Algorithms and Software* and *Cognitive Systems*.

## 8 LESSONS LEARNED

The notable lesson we can conclude is that what shapes the cognitive computation is the difference between the old and new logic and the old and new hardware. Moreover, Cognitive computing is extending of being a brain-like-computing paradigm to a strategic technology. In the following sections, we are discussing the findings in each of the new computing paradigms

### 8.1 Neuromorphic Applications

Neuromorphic chips are used primarily for building spike neural networks. Their physical neural network structure offers low power with extremely efficient parallel computing and short training times. While GPUs accelerated deep learning training and testing, they lack the physical structure of neurons which dissimilarities cause a lot of inefficiencies, such as excessive power consumption. The speed of communication between synapses in neuromorphic chips reach real time performance and solve the delay in applications that require real-time response such as electric vehicle [5, 126]. However, NE circuits are used now as artificial intelligence accelerators and the artificial intelligence accelerators and co-processors rather than independent Turing machines.

Neuromorphic chips such as Tianjic [127], allow multiple cognitive tasks such as speech recognition and object detection concurrently and using one chip. The low power consumption and simple structure of neural networks in neuromorphic chips, will open new venues for internet of things and robotic applications to run cognitive tasks at real time with modest consumption of resources. NE will be used primarily for simulations, optimization, and graph algorithms tasks to solve NP-hard problems. Moreover, using neuromorphic chips with IoTs will provide efficient and simple computation [119]

### 8.3 Open Research Issues

In this section, we are presenting long term vision on future research and challenges.

***Scalability:*** Many if not virtually all current use of quantum computing is proof of concept projects or research. However, neuromorphic chips are not produced at scale that allows applying them in industrial and domestic applications. QC remains out of reach for classical computers for the foreseeable future: Large-scale, combinatorics calculations.

Most of research on quantum computing is purely theoretical. So, quantum algorithms that tackle real-world problems cannot run on a scale because of the limited development of efficient quantum hardware to apply quantum algorithms. QC and NE algorithms must move from the narrow scientific application to commercial applications. Despite the notable development in QC algorithms such as Grover's algorithm, QC is illsuited for large-scale search applications, such as querying of databases [128]. On the other hand, while NE hardware is inherently scalable, they need massive scalability (millions of neurons) to perform reliably.

***Decoherence:*** QC requires complicated hardware special and environmental condition to generate quantum states (entanglement and coherence). So, photons/ neurons are sensitive to noise and lose their coherence state with slight change in the environment. This issue is one of the main challenges against scaling QC for business and industrial use. Generating with Quantum hardware that could operate in room temperature and resistant to noise and environment is going to create a leap in Cognitive Computation.

**Programming languages and libraries:** the lack of programming languages for QC an NE paradigms is limiting the development of cognitive systems beyond the research purposes. Moreover, the lack of programming libraries that facilitate the communication between classical computing and other paradigms are still wide areas of research.

***Quantum-Safe Encryption:*** Until a Quantum circuit with thousands of qubits is available, most of current encryption systems would be compromised. This calls for immediate research on post-quantum cryptography methods that is Quantum-Safe or cannot be compromised easily using Quantum machines. Lattice-based cryptography and Hash-based cryptography are two actively research approaches [129].

***Versatile AI***: There are a computer-science-oriented, Quantum-oriented and a neuroscience-oriented approaches for developing AI. There are fundamental differences in their formulations and coding schemes, and they rely on incompatible platforms. Developing hybrid synergistic platforms that combine the three approaches and integrate different AI formulations orientation to allow simultaneous processing of versatile algorithms and models is another open research issue.

***Strategic Cognitive Computing:*** cognitive computing specially QC has potential to change the cryptography and decision-making games. The strategic dimensions of CC are strongly tight to security, productivity, operations management, and proactive innovation. So, CC has many strategic dimensions, and the development of new CC hardware/software is creating competitive advantage for countries both outwardly and inwardly.

## 9 CONCLUSION

Defining what is cognitive computing has been a subject of deliberation. Cognitive Computing is the emergent transformation in computing paradigms that aims to transfer human cognition to machines. Human cognition entails three different computing paradigms (Von-Neumann, Neuromorphic computing, and Quantum computing). However, none of them could explain human cognition fully but together they orchestrate different complex cognitive tasks. We conducted a systematic literature review to trace the structure and state of art in Cognitive Computing research. We used different statistical analysis techniques to analyze the CC literature that resulted in a pyramid-like architecture with four sections of *Representation* *of intelligence, Brain-like Hardware, Cognitive Algorithms and Software, and Cognitive Systems*. In the last decade, significant advancements in *AI* and *Brain-like Hardware* fields have catapulted CC development. However, the development in NE and Quantum hardware has a long way to go to ultimately complement the CC architecture. One of the gaps in defining CC that NE and quantum hardware and software research don't consider CC as an umbrella of development or an important goal to deem. The research summarized the lessons learned from the literature revies and suggested future research issues.

## 10 REFERENCES

[1] Sreedevi, A., Harshitha, T. N., Sugumaran, V. and Shankar, P. Application of cognitive computing in healthcare, cybersecurity, big data and IoT: A literature review. *Information Processing & Management*, 59, 2 (2022), 102888.

[2] Wang, Y. Cognitive robots. *IEEE robotics & automation magazine*, 17, 4 (2010), 54-62.

[3] Wang, Y., Zhang, D. and Tsumoto, S. Preface: Cognitive Informatics, Cognitive Computing, and Their Denotational Mathematical Foundations (I). *Fundamenta Informaticae*, 90, 3 (2009), i-vii.

[4] Pescovitz, D. Helping computers help themselves. *IEEE Spectrum*, 39, 9 (2002), 49-53.

[5] Wang, W., Deng, X., Ding, L. and Zhang, L. *Neural Cognitive Computing Mechanisms*. Springer, City, 2020.

[6] Indiveri, G., Chicca, E. and Douglas, R. J. Artificial cognitive systems: From VLSI networks of spiking neurons to neuromorphic cognition. *Cognitive Computation*, 1, 2 (2009), 119-127.

[7] Schinazi, R. B. *Combinatorics*. Springer, City, 2022.

[8] Sinz, F. H., Pitkow, X., Reimer, J., Bethge, M. and Tolias, A. S. Engineering a less artificial intelligence. *Neuron*, 103, 6 (2019), 967-979.

[9] Semantic-Web, C. Introducing Semantic AI, Ingredients for a sustainable Enterprise AI Strategy. *White paper* (2019).

[10] Hrnjica, B. and Mehr, A. D. *Energy Demand Forecasting Using Deep Learning*. Springer, City, 2020.

[11] Chen, Z. and Yang, Z. Graph Neural Reasoning May Fail in Proving Boolean Unsatisfiability. *arXiv preprint arXiv:1909.11588* (2019).

[12] Russell, J. and Cohn, R. Moravec's Paradox. *Book on Demand*, 15 (2012).

[13] Wang, Y. On abstract intelligence and brain informatics: Mapping cognitive functions of the brain onto its neural structures. *International Journal of Cognitive Informatics and Natural Intelligence (IJCINI)*, 6, 4 (2012), 54-80.

[14] Shah, H. *Turing's misunderstood imitation game and IBM's Watson success*. City, 2011.

[15] Schuetz, S. and Venkatesh, V. Research Perspectives: The Rise of Human Machines: How Cognitive Computing Systems Challenge Assumptions of User-System Interaction. *Journal of the Association for Information Systems*, 21, 2 (2020), 2.

[16] Chen, M., Herrera, F. and Hwang, K. Cognitive Computing: Architecture, Technologies and Intelligent Applications. *IEEE Access*, 6 (2018), 19774-19783.

[17] Chen, M., Hao, Y., Gharavi, H. and Leung, V. C. Cognitive information measurements: A new perspective. *Information Sciences*, 505 (2019), 487-497.

[18] Ray, A., Bala, P. K. and Dasgupta, S. A. Role of authenticity and perceived benefits of online courses on technology based career choice in India: A modified technology adoption model based on career theory. *International Journal of Information Management*, 47 (2019), 140-Newell, A. *Unified theories of cognition*. Harvard University Press, 1994.

[21] Pinker, S. How the mind works. *Annals of the New York Academy of Sciences*, 882, 1 (1999), 119-127.

[22] Elnagar, S. and Thomas, M. *Explaining Cognitive Computing Through the Information Systems Lens*. AIS, City, 2020.

[23] Wisse, P. The Constitutional Force of Perspective Phenomenology: Philosophical Unification in Information Systems. *AMCIS 2003 Proceedings* (2003), 363.

[24] Nickerson, R. C., Varshney, U. and Muntermann, J. A method for taxonomy development and its application in information systems. *European Journal of Information Systems*, 22, 3 (2013), 336-359.

[25] Goguen, J. A. and Linde, C. *Techniques for requirements elicitation*. IEEE, City, 1993.

[50] Ahmed, H., Intaravanne, Y., Ming, Y., Ansari, M. A., Buller, G. S., Zentgraf, T. and Chen, X. Multichannel superposition of grafted perfect vortex beams. *Advanced Materials* (2022), 2203044.

[51] Stratton, P. Convolutionary, Evolutionary, Revolutionary: What's next for Bodies, Brains and AI? (2019).

[52] Miller, S. AI: Augmentation, more so than automation (2018).

[53] Wang, Y. *Basic theories for neuroinformatics and neurocomputing*. IEEE, City, 2013.

[54] Wang, Y. In search of denotational mathematics: Novel mathematical means for contemporary intelligence, brain, and knowledge sciences. *Journal of Advanced Mathematics and Applications*, 1, 1 (2012), 4-26.

[55] Moschovakis, Y. N. Sense and denotation as algorithm and value. *Lecture notes in logic*, 2 (1994), 210-249.

[56] Kunii, T. L. *Autonomic and trusted computing for ubiquitous intelligence*. Springer, City, 2007.

[57] Wang, Y., Zadeh, L. A., Widrow, B., Howard, N., Beaufays, F., Baciu, G., Hsu, D. F., Luo, G., Mizoguchi, F. and Patel, S. *Abstract intelligence: Embodying and enabling cognitive systems by mathematical engineering*. IGI Global, City, 2020.

[58] Taplin, W. L. Six general principles of intelligence. *International Journal of Intelligence and Counter Intelligence*, 3, 4 (1989), 475-491. [71] Wang, Y., Wang, Y., Patel, S. and Patel, D. A layered reference model of the brain (LRMB). *IEEE Transactions on Systems, Man, and Cybernetics, Part C (Applications and Reviews)*, 36, 2 (2006), 124-133.

[72] Khrennikov, A. Y. and Haven, E. Quantum mechanics and violations of the sure-thing principle: The use of probability interference and other concepts. *Journal of Mathematical Psychology*, 53, 5 (2009), 378-388.

[73] Song, Q., Wang, W., Fu, W., Sun, Y., Wang, D. and Gao, Z. Research on quantum cognition in autonomous driving. *Scientific reports*, 12, 1 (2022), 1-16.

[91] Mordvintsev, A., Olah, C. and Tyka, M. Inceptionism: Going deeper into neural networks (2015).

[92] Lobo, J. L., Del Ser, J., Bifet, A. and Kasabov, N. Spiking neural networks and online learning: An overview and perspectives. *Neural Networks*, 121 (2020), 88-100.

[93] Pei, J., Deng, L., Song, S., Zhao, M., Zhang, Y., Wu, S., Wang, G., Zou, Z., Wu, Z. and He, W. Towards artificial general intelligence with hybrid Tianjic chip architecture. *Nature*, 572, 7767 (2019), 106-111.

[94] Ang, C. H., Jin, C., Leong, P. H. and van Schaik, A. *Spiking neural network-based auto-associative memory using FPGA interconnect delays*. IEEE, City, 2011.

[95] Esser, S. K., Appuswamy, R., Merolla, P., Arthur, J. V. and Modha, D. S. *Backpropagation for energy-efficient neuromorphic computing*. City, 2015.

[96] Hill, A. *SNL Fugu to IBM's TrueNorth Corelet Programming Environment Transcompiler*. Sandia National Lab.(SNL-NM), Albuquerque, NM (United States), 2020.

[97] Jin, H. Data processing model and performance analysis of cognitive computing based on machine learning in Internet environment. *Soft Computing*, 23, 19 (2019), 9141-9151.

[98] Amir, A., Datta, P., Risk, W. P., Cassidy, A. S., Kusnitz, J. A., Esser, S. K., Andreopoulos, A., Wong, T. M., Flickner, M. and AlvarezIcaza, R. *Cognitive computing programming paradigm: a corelet language for composing networks of neurosynaptic cores*. IEEE, City, 2013.

[99] Davies, M. Benchmarks for progress in neuromorphic computing. *Nature Machine Intelligence*, 1, 9 (2019), 386-388.

[100] Pronin, C. B. and Ostroukh, A. V. Developing Mathematical Oracle Functions for Grover Quantum Search Algorithm. *arXiv preprint arXiv:2109.05921* (2021).

[101] Xie, H. and Yang, L. A quantum related-key attack based on the Bernstein–Vazirani algorithm. *Quantum Information Processing*, 19, 8 (2020), 1-20.

[102] Combarro, E. F., Piñera-Nicolás, A., Ranilla, J. and Rúa, I. F. An explanation of the Bernstein-Vazirani and Deustch-Josza algorithms with